\documentclass[runningheads]{llncs}
\usepackage[T1]{fontenc}
\usepackage{graphicx}
\usepackage{amsmath}
\usepackage{float}

\begin{document}
\title{ACTISM: Threat-Informed Dynamic Security Modelling for Automotive Systems}
\titlerunning{Threat-Informed Dynamic Security Modelling for Automotive Systems}
%
\author{Shaofei Huang \and
Christopher M. Poskitt \and
Lwin Khin Shar}
\authorrunning{S. Huang et al.}
%
\institute{Singapore Management University, Singapore\\
\email{\{sf.huang.2023,cposkitt,lkshar\}@smu.edu.sg}}
\maketitle              
\begin{abstract}
Evolving cybersecurity threats in complex cyber-physical systems pose significant risks to system functionality and safety. This experience report introduces ACTISM (Automotive Consequence-Driven and Threat-Informed Security Modelling), an integrated security modelling framework that enhances the resilience of automotive systems by dynamically updating their cybersecurity posture in response to prevailing and evolving threats, attacker tactics, and their impact on system functionality and safety. ACTISM addresses the existing knowledge gap in static security assessment methodologies by providing a dynamic and iterative framework. We demonstrate the effectiveness of ACTISM by applying it to a real-world example of the Tesla Electric Vehicle's In-Vehicle Infotainment system, illustrating how the security model can be adapted as new threats emerge. We also report the results of a practitioners' survey on the usefulness of ACTISM and its future directions. The survey highlights avenues for future research and development in this area, including automated vulnerability management workflows for automotive systems.

\keywords{Risk management  \and Security \and Automotive systems}
\end{abstract}
\section{Introduction}
\label{sec:introduction}

Automotive systems have evolved into sophisticated cyber-physical systems, packed with cutting-edge technologies. Unlike their predecessors, today’s vehicles are equipped with advanced sensors, including LiDAR (Light Detection and Ranging), radar, and cameras, alongside a variety of connectivity options such as Control Area Networks (CAN), Local Interconnect Networks (LIN), FlexRay, and Ethernet. External connectivity is further enhanced through Bluetooth, Wi-Fi, and GPS, enabling seamless integration with external devices and services. Additionally, the integration of feature-rich In-Vehicle Infotainment (IVI) systems and powerful computing units for autonomous driving continues to push the boundaries of automotive innovation. However, this technological evolution significantly broadens the attack surface, making automotive systems increasingly vulnerable to cybersecurity threats.

Cybersecurity threats and vulnerabilities in the automotive industry are well-documented, with attackers exploiting these weaknesses to compromise vehicle security, access sensitive data, and even take control of critical functions.

Recent literature underscores the growing challenges of defending automotive systems against such threats. For instance, Pham and Xiong \cite{Pham21}, in their survey on security attacks and defence techniques for Connected and Autonomous Vehicles (CAVs), highlighted how attackers can sequentially compromise vehicle components. They stressed the importance of securing all system elements to ensure comprehensive protection. While this approach is viable during the early stages of a vehicle's life cycle, it becomes less effective over time as cybersecurity threats evolve in both complexity and scale. The rapid advancement of automotive technologies further exacerbates this issue, making it impractical to rely solely on initial cybersecurity measures to address emerging risks throughout a vehicle's operational life.

This is where \emph{dynamic threat and attack models} prove invaluable. By providing a proactive and adaptive framework, these models enable engineers to continuously analyse and mitigate cybersecurity risks, ensuring systems remain resilient against evolving threats. Additionally, they empower consumers to make informed decisions by evaluating a vehicle's cybersecurity robustness.
However, many existing threat and attack models do not fully address the unique cybersecurity requirements of modern automotive systems. These requirements include:

\begin{itemize}

    \item Interconnectivity: Modern vehicles are highly interconnected systems, comprising numerous Electronic Control Units (ECUs) that communicate with each other. Ensuring secure communication between these components is essential to prevent unauthorised access and control.

    \item Safety-Critical Systems: The integration of advanced driver-assistance systems (ADAS) and autonomous driving technologies means that any cybersecurity breach could directly impact vehicle safety. Ensuring the security of these systems is paramount to protecting the safety of passengers and other road users.

    \item Physical Security: Automotive systems are subject to physical threats. Attackers can potentially gain physical access to vehicles, enabling them to tamper with hardware components or inject malicious code directly.

    \item Over-the-Air (OTA) Updates: Many modern vehicles support OTA updates for software and firmware. While this feature enhances convenience and functionality, it also introduces new attack vectors.

    \item User Privacy: Automotive systems collect and process a significant amount of data, including personal information, driving habits, and location data. Protecting user privacy is crucial to prevent misuse of this data.
    
\end{itemize}

In our previous work~\cite{Huang24SLR}, we conducted a systematic literature review of threat and attack modelling approaches pertinent to cyber-physical systems~(CPSs)---systems that integrate computational and physical process---of which automotive systems are an example. Our investigation yielded several findings, including:

\begin{itemize}

    \item Threat modelling is commonly conducted only in early stages of system development, and does not always consider attack models.

    \item Threat and attack models in the literature primarily focus on IT systems, making them challenging to apply to automotive systems.

    \item Much of the existing literature does not differentiate between cybersecurity breaches and consequences in IT systems versus those in automotive systems.

    \item Ambiguity persists in the literature regarding the definitions and relationships between threat modelling and attack modelling.
    
\end{itemize}

For instance, threat models developed using Microsoft’s STRIDE (Spoofing, Tampering, Repudiation, Information Disclosure, Denial of Service, and Elevation of Privilege) primarily focus on data flows and information security. 
As a result, current threat modelling practices for automotive systems often fail to consider the broader consequences of cybersecurity attacks beyond information security \cite{Durrwang18}. They often overlook critical aspects of how cybersecurity breaches can directly impact automotive functions and safety, raising questions about the reliability of their outcomes when applied to the unique characteristics of automotive environments. Similarly, attack models designed for traditional IT systems do not fully account for the multi-path tactics and techniques attackers use to gain control of a vehicle \cite{Chlup22}\cite{Ebrahimi22}. These techniques often involve a \emph{combination of physical and cyber actions}, highlighting the need for a more comprehensive and dynamic approach to security modelling.

These findings underscore the importance of tailoring security modelling approaches to automotive systems, ensuring they adequately capture the unique cybersecurity threats these environments face.

\textit{Contributions}. In this experience report, we introduce ACTISM (Automotive Consequence-Driven and Threat-Informed Security Modelling) — an integrated security modelling framework that enhances the resilience of automotive systems by \emph{dynamically updating their cybersecurity posture} in response to prevailing and evolving threats, attacker tactics, and their impact on system functionality and safety. This approach addresses the existing knowledge gap in static security assessment methodologies by providing a dynamic and iterative framework. We demonstrate the effectiveness of ACTISM by applying it to an IVI system, showing how the model can be adapted as new threats emerge.

Unlike existing security modelling frameworks, ACTISM \emph{integrates multiple methodologies} to dynamically update security models with new information gathered through threat intelligence and continuous security monitoring. This ensures that the models remain relevant against evolving attacker tactics, techniques, and procedures (TTPs), even if they were initially developed during the early stages of system design. ACTISM also enables practitioners to apply these security models across both IT and automotive domains by utilising attack paths to accurately represent the multi-path and multi-agent nature of real-world cyber-physical attacks. To enhance the comprehensiveness of its security model, ACTISM applies and adapts the following threat analysis and risk assessment frameworks:

\begin{itemize}

    \item STRIDE, a widely-used threat modelling framework developed by Microsoft to characterize security threats in software systems. The acronym STRIDE stands for six types of threats: Spoofing, Tampering, Repudiation, Information Disclosure, Denial of Service, and Elevation of Privilege. 
    In addition to information security, ACTISM reports how physical threats can be modeled using STRIDE in conjunction with other frameworks (Section~\ref{sec:application}).

    \item HEAVENS (i.e., HEAling Vulnerabilities to ENhance Software Security and Safety), a structured and model-based hazard analysis and risk assessment method specifically designed for automotive systems.
    
    \item Attack Tree, a hierarchical diagram representing potential attack paths, used for modeling how an attacker could exploit system vulnerabilities to achieve specific goals.

    \item Common Vulnerability Scoring System (CVSS), an open industry standard for assessing the severity of computer system security vulnerabilities. This model is typically used in a static manner, i.e., the severity scores are usually not updated for new or evolving threats. ACTISM shows how CVSS can be applied dynamically (Section~\ref{sec:updating}).  

    \item Threat Analysis and Risk Assessment (TARA) methodology adopted from the ISO/SAE 21434:2021 ("Road Vehicles - Cybersecurity Engineering") Standard, jointly developed by the International Organisation for Standardisation (ISO) and the Society of Automotive Engineers (SAE), for managing cybersecurity threats and vulnerabilities throughout the vehicle's life cycle.
    
\end{itemize}

While frameworks like STRIDE and CVSS were originally designed for IT security, they can also be applied to physical threats. For instance, tampering with physical components can lead to unauthorised access, data manipulation, or service disruption. ACTISM takes an integrative approach, combining these frameworks to model both cyber and physical attacks. This ensures comprehensive coverage of potential entry points and attack methods, addressing the multi-faceted nature of modern automotive threats.

To demonstrate ACTISM's practical application, we developed a detailed security model for an In-Vehicle Infotainment (IVI) system. The step-by-step process highlights the integration of threat intelligence and contextual information to accurately characterise and assess cybersecurity risks. Additionally, we illustrate ACTISM's adaptive capabilities by simulating its response to a real-world vulnerability disclosure affecting the IVI system. This case study underscores ACTISM's effectiveness in addressing emerging threats and evolving security challenges, while also reporting its versatility for application across other automotive systems beyond IVI.

Finally, we report the results of a practitioners' survey on the usefulness of ACTISM and its future directions. The survey highlights avenues for future research and development in this area, including automated vulnerability management workflows for automotive systems. These advancements aim to further enhance ACTISM's ability to safeguard automotive systems against increasingly sophisticated cybersecurity threats.

\section{Related Work}


Das et al.~\cite{Das24} employed the STRIDE methodology for threat modelling and utilised the SAHARA (Safety-Aware Hazard Analysis and Risk Assessment) and DREAD (Damage, Reproducibility, Exploitability, Affected Users, and Discoverability) methodologies to assess risks associated with Infotainment High Performance Computing (HPC) systems. However, their analysis did not incorporate attack modelling. 

Abuabed et al. \cite{Abuabed23} proposed a framework for assessing the cybersecurity vulnerabilities of automotive vehicles. Their approach employed STRIDE to classify threats. For assessing the impact of threats, they used version 2.0 of the HEAVENS security model \cite{Lautenbach21}, and they evaluated attack feasibility with CVSS version 2.0. Additionally, they used a risk matrix to assess risks based on a use-case involving the Advanced Driver Assistance System (ADAS). Using a TARA approach compatible with the ISO/SAE 21434:2021 standard, the framework utilises attack trees to determine attack paths. This method facilitates a focused analysis of relevant threat vectors and potential attacks during the assessment of attack feasibility.

However, the framework has several limitations. Firstly, in the threat identification process, the authors did not include physical attack vectors, such as accessing and modifying embedded software in vehicle components, in the STRIDE Data Flow Diagram (DFD). This omission could be because of the assumption that physical attack vectors do not involve flows or exchanges of data. Notwithstanding this assumption, physical attack vectors can be exploited to facilitate data flows, such as "USB Man-in-the-Middle (USB MITM)" attacks, where an attacker intercepts and potentially alters the communication between a USB device and a host system. Consequently, physical attack vectors can be a crucial gap in the overall security design of the vehicle.
Secondly, the CVSS version 2.0 framework adopted in Abuabed et al. \cite{Abuabed23} lacks the granularity and accuracy that newer versions of CVSS can provide, which impacts the resulting risk assessment. Lastly, the framework does not adequately address threats in later stages of the life cycle. As automotive system life cycles are typically 10 to 20 years or longer, the vehicle system could become increasingly vulnerable to new types of attacks that were not anticipated during the initial security assessment.

Other literature relevant to automotive cybersecurity includes a study by Girdhar et al.~\cite{Girdhar22} who used the STRIDE methodology to create a threat model focused on post-accident cyberattack event analysis for CAV. Chlup et al.~\cite{Chlup22} proposed an automated threat modelling tool for the automotive domain. It combines threat modelling and attack tree analysis using automatically-generated attack trees and attack paths. In addition, it uses anti-patterns to perform in-depth analysis in the threat model. However, the approach does not perform attack feasibility estimation and risk assessment, which are key steps in the TARA workflow. Dang et al.~\cite{Dang20}  used a STRIDE framework to analyse and classify cybersecurity threats in an autonomous vehicle architecture, including the infotainment system. While the authors ranked the threats using CVSS version 3.0 scores, they did not assess the corresponding impact. The authors focused only on cybersecurity threats originating from outside the vehicle. This means that attacks that involve access to the vehicle's internal architecture from within the vehicle are excluded in the analysis.

In summary, the current landscape of security modelling highlights several limitations. Firstly, there are differing assumptions regarding which vehicle assets fall within the trust boundaries of security models. Secondly, the feasibility and impact of cyber-physical attacks are rarely considered in security modelling. These gaps underscore the need for an iterative and cohesive framework to effectively address the evolving challenges in automotive cybersecurity.

\section{High-Level Overview of ACTISM}

We propose ACTISM, a framework for modelling security threats and attack vectors in a dynamic, iterative manner, specifically designed for automotive systems. Unlike static models, ACTISM incorporates threat intelligence through continuous monitoring, which ensures that the security model remains adaptive and effective in mitigating emerging risks, like those targeting keyless entry systems \cite{Wouters21},  breaching the trust boundary within security models.

\begin{figure}[t]
    \centering
    \includegraphics[width=0.8\textwidth]{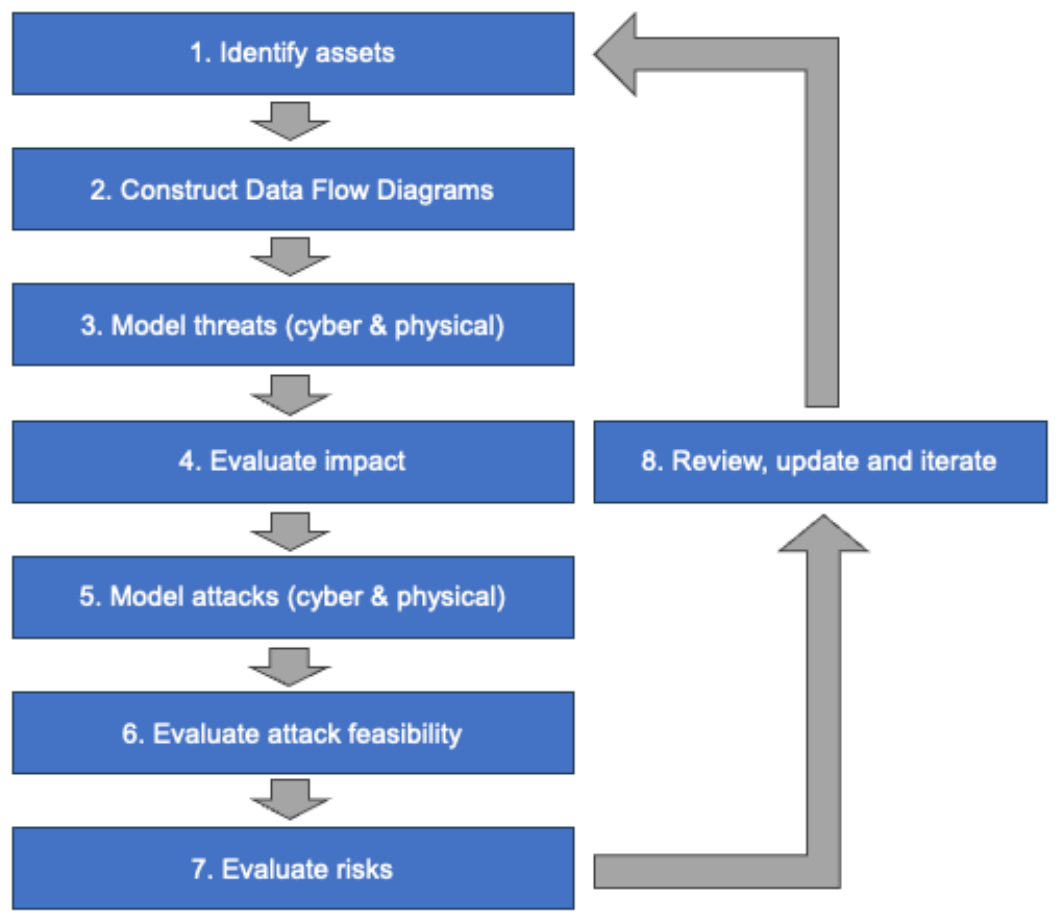}
    \caption{Overview of ACTISM}
    \label{fig:ACTISM_Overview}
\end{figure}

ACTISM is not designed to replace existing threat and attack modelling frameworks. Rather, it simplifies their adoption and application for cybersecurity practitioners. By integrating multiple security modelling frameworks into a cohesive, iterative process, ACTISM significantly improves the practicality and effectiveness of addressing evolving cybersecurity threats and attacker capabilities.

As illustrated in Figure \ref{fig:ACTISM_Overview}, ACTISM consists of eight key steps. These steps integrate multiple other threat and risk assessment frameworks, such as STRIDE, HEAVENS, attack trees, CVSS, and TARA, as outlined in Section \ref{sec:introduction}. The details of each step will be expanded upon in Section \ref{sec:application}.


To enhance the practical utility of ACTISM among cybersecurity practitioners, we utilise HEAVENS version 2.0 \cite{Lautenbach21} for impact assessment. To our knowledge, HEAVENS is the only framework designed to be fully compliant with the latest automotive cybersecurity standards, such as ISO/SAE 21434 and UNECE Regulation 155.
In addition, we employ CVSS version 3.1 to assess attack feasibility, which allows us to leverage contemporary cybersecurity vulnerability knowledge bases. CVSS version 3.1 provides granular metrics and scoring options, considering various factors that affect the exploitability and impact of a vulnerability. This comprehensive approach ensures that ACTISM remains robust and adaptive to the dynamic nature of cybersecurity threats in automotive systems.

\section{Applying ACTISM to an IVI System}
\label{sec:application}

To demonstrate ACTISM's utility, we use an In-Vehicle Infotainment (IVI) system as a running example and apply it to a real-world case study. To appreciate the significance of this application, it is essential to understand the role of IVI systems in modern vehicles and their associated cybersecurity challenges.

Legacy head units offered basic functionalities such as radio reception and CD playback, with minimal connectivity. In contrast, modern IVI systems are sophisticated digital platforms that integrate a wide range of features beyond audio and navigation. They act as central hubs, managing complex data streams, enabling access to vehicle diagnostics, and providing interactive user interfaces for drivers and passengers. Additionally, they connect to external networks, mobile devices, and cloud-based services, significantly enhancing convenience and operational efficiency.

However, these advancements also introduce substantial cybersecurity risks. Modern IVI systems are vulnerable to traditional threats like remote exploits and data breaches, as well as emerging cyber-physical attacks that target vehicle functionalities. The integration of multiple wireless communication protocols such as Wi-Fi, Bluetooth, and cellular networks expands the attack surface, providing malicious actors with opportunities to exploit weaknesses in wireless protocols or insecure network connections. Advanced features like voice recognition and artificial intelligence further increase complexity, creating additional attack vectors.

A dynamic framework like ACTISM is uniquely suited to address these challenges. By enabling iterative security modelling throughout the IVI system's life cycle, ACTISM ensures comprehensive protection against evolving threats, safeguarding both functionality and safety in modern vehicles.

\subsection{Step 1: Identify assets}

In the first step, identifying vehicle assets within the IVI system threat model typically involves a collaborative effort among system architects, engineers, and cybersecurity experts. This process integrates various activities, including reviewing technical specifications, examining system architecture diagrams, and analysing functional requirements documents. Additionally, considering the interconnected nature of modern IVI systems, potential dependencies and interactions between assets are also analysed to understand how vulnerabilities in one component could impact overall system security. These efforts aim to ensure a comprehensive understanding and coverage of all relevant assets within the IVI system.

In this experience report, we build on the IVI HPC system DFD from Das et al. (2024) \cite{Das24} to guide our identification of IVI system assets. This initial asset identification, listed below, serves as a foundational step for conducting further detailed analysis and security modelling. While the referenced DFD provides a comprehensive framework for understanding IVI system components and interactions, we extend its scope to include emerging threats, such as dirty road attacks \cite{sato21}. These attacks exploit the fact that autonomous and semi-autonomous vehicles rely heavily on sensors and cameras to interpret their environment. Attackers may place malicious objects, patterns, or signals on the road to confuse these sensors, leading to incorrect navigation decisions, system malfunctions, or even accidents. By incorporating such attack vectors, we aim to enhance the  cyber-physical security modelling process and address the evolving threat landscape faced by modern automotives.

\begin{itemize}
    \item \textbf{In-vehicle modules}
    \begin{itemize}
        \item In-Vehicle Infotainment (IVI) System
        \item Human-machine Interface (HMI)
        \item Central Gateway
        \item GPS Receiver
        \item USB
        \item Bluetooth
        \item Wi-Fi
    \end{itemize}

    \item \textbf{Data stores}
    \begin{itemize}
        \item Local maps data store
        \item IVI system embedded data store
    \end{itemize}
\end{itemize}

\subsection{Step 2: Construct Data Flow Diagrams}

\begin{figure*}[t]
    \centering
    \includegraphics[width=1.0\textwidth]{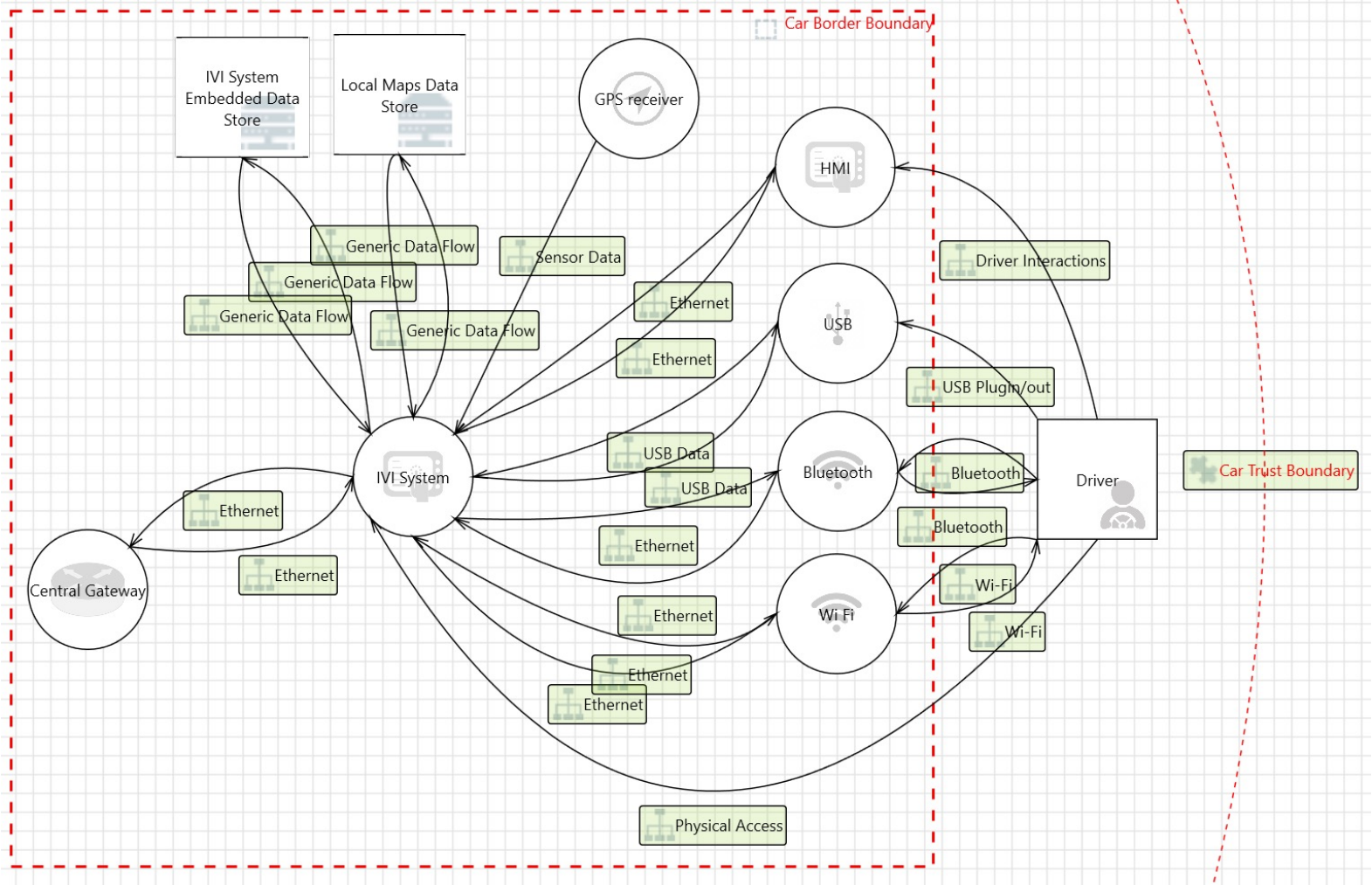}
    \caption{Data Flow Diagram for an In-Vehicle Infotainment (IVI) System}
    \label{fig:DFD_IVI}
\end{figure*}

In the next step, we construct and analyse DFDs to identify the vehicle trust boundary and potential threats  and attack vectors, taking into consideration physical aspects. This process uses the Microsoft Threat Modelling Tool (Version 7.3.31026.3) along with the Automotive Threat Modelling Template \cite{NCC} developed by the NCC Group. We conduct DFD analysis for both a typical automotive system \cite{Huang24A} and specifically for the IVI system (Figure \ref{fig:DFD_IVI}) to ensure thorough identification and analysis of potential threats and vulnerabilities.

\subsection{Step 3: Model threats (cyber \& physical)}
Next, we use the Microsoft Threat Modelling Tool to generate threats based on the previously constructed Data Flow Diagram (DFD) for the IVI system. The tool categorises these threats according to the STRIDE threat model automatically, providing an initial classification into categories such as Spoofing, Tampering, Repudiation, Information Disclosure, Denial of Service, and Elevation of Privilege. In addition to cyber threats, we also model physical threats, such as unauthorised physical access to the IVI system or tampering with hardware components.

While the tool categorises threats by default and attempts to provide threat descriptions based on its own threat catalogues and templates, the initial output may be somewhat generic and not specifically tailored to the nuances of the IVI system's operating environment. Therefore, our analysis involves further contextualisation and systematic review of these categorised threats.

We focus on the in-vehicle modules and data stores identified in the earlier step, ensuring that the generated threats are relevant to these components. After refining the list, we update threat descriptions to better align with the specific interfaces and nuances of the IVI system. A domain expert can perform this refinement relatively easily using the Microsoft Threat Modelling Tool, taking about 30 minutes for this example. The final curated list of threats for the IVI system is shown in Table \ref{Table_Threats} of our online appendix (see \cite{Huang25A}).

\subsection{Step 4: Evaluate impact}

To evaluate the impact of the identified threats, we use the HEAVENS version 2.0 framework \cite{Lautenbach21} which is specifically designed for automotive risk assessments.

Our approach to impact analysis is as follows:

\begin{itemize}
    \item \textbf{Assess impact level}:
    By definition in the HEAVENS framework, the impact level of a specific threat is an estimate of the expected loss for various stakeholders when the threat is realised. To estimate the impact, we use four parameters directly related to the security objectives defined in HEAVENS: safety (s), financial (f), operational (o), and privacy (p).
    
    \begin{itemize}
        \item Safety Impact: Refers to the safety of vehicle occupants, road users, and infrastructure.
        \item Financial Impact: Encompasses all direct and indirect financial damages incurred by stakeholders.
        \item Operational Impact: Pertains to operational damages that have minimal or no safety or financial consequences. For example, the loss of secondary functionalities such as entertainment systems.
        \item Privacy Impact: Deals with damages resulting from privacy violations.
    \end{itemize}
    
    Each of the four parameters is assigned an impact level: None, Low, Medium, or High, based on definitions from \cite{Lautenbach21}. These levels are typically determined by automotive domain experts with in-depth knowledge of the system and its vulnerabilities. For this report, we assigned impact levels based on our assessment of potential consequences, informed by literature, though these may differ from expert evaluations.
    
    To quantify these levels, numerical values are assigned on a logarithmic scale (0, 1, 10, 100), reflecting the exponential increase in impact severity. This approach ensures clear differentiation between impact levels and prioritises high-impact threats effectively. Table \ref{Table_Impact_Assessment} \cite{Huang25A} categorises the impact levels and parameter values according to the security objectives.
    
    \item \textbf{Calculate impact score}:
    After estimating the individual impact levels of the security objectives and their corresponding parameter values, the overall impact score is computed using the following equation:
    
    \begin{equation}
    I_{\text{sum}} = w_{s} i_{s} + w_{o} i_{o} + w_{f} i_{f} + w_{p} i_{p}
    \label{eq:Equation_Impact_Sum}
    \end{equation}
    
    Normalisation is necessary to ensure that the overall impact score is on a consistent scale, allowing for fair and meaningful comparisons across different systems and contexts. This uniformity is crucial when comparing the cybersecurity posture of various systems, such as between automotive systems.
    
    The overall normalised impact score is computed using the following equation:

    \begin{equation}
        I_{\text{nsum}} = \frac{w_{s} i_{s} + w_{o} i_{o} + w_{f} i_{f} + w_{p} i_{p}}{100 * (w_{s} + w_{o} + w_{f} + w_{p})}
        \label{eq:Equation_Impact_Sum_Normalised}
    \end{equation}
    
    Weight values \(w_{s}\), \(w_{o}\), \(w_{f}\), and \(w_{p}\) are selected based on the specific needs of the use case being evaluated. In the case of an automotive system, the safety impact is the most severe consequence for stakeholders, such as vehicle occupants, road users, and infrastructure. Comparatively, financial, operational, and privacy impacts may have lower consequences. As such, the weights are selected as: \(w_{s} = 10\) and \(w_{o} = w_{f} = w_{p} = 1\). Equations \ref{eq:Equation_Impact_Sum} and \ref{eq:Equation_Impact_Sum_Normalised} are simplified as follows:

    \begin{equation}
        I_{\text{sum}} = 10i_{s} + i_{o} + i_{f} + i_{p}
        \label{eq:Equation_Weighted_Impact_Sum}
    \end{equation}

    \begin{equation}
        I_{\text{nsum}} =  \frac{10i_{s} + i_{o} + i_{f} + i_{p}}{1300}
        \label{eq:Equation_Weighted_Impact_Sum_Normalised}
    \end{equation}

    \item \textbf{Assign impact rating}:
    Each threat is assigned an impact rating based on the impact score \(I_{\text{nsum}}\), matched against thresholds tabulated in Table \ref{Table_Impact_Rating_Mapping}, as referenced from HEAVENS 2.0 \cite{Lautenbach21}. The unequal ranges of the impact scores emphasise the importance of distinguishing between high-severity and critical-severity impacts, particularly in safety-critical systems like automotive systems. Similarly, the smaller ranges for lower-severity impacts reflect the fact that these scenarios are less critical and require less granularity in their assessment.

    \begin{table}[H]
        \centering
        \caption{Impact rating table}
        \label{Table_Impact_Rating_Mapping}
        \begin{tabular}{|c|c|}
            \hline
            \textbf{Impact Score ($I_{\text{nsum}}$)} & \textbf{Impact Rating} \\
            \hline
            0.00 & None \\
            $0.00 < x < 0.01$ & Low \\
            $0.01 \leq x < 0.05$ & Medium \\
            $0.05 \leq x < 0.45$ & High \\
            $0.45 \leq x < 1.00$ & Critical \\
            \hline
        \end{tabular}
         
    \end{table}

\end{itemize}

For illustration, we show how the impact rating of Threat ID 9146 from Table \ref{Table_Threats} \cite{Huang25A} is derived using this methodology. In this specific threat, an adversary may reverse engineer the head unit firmware to find sensitive information, resulting in information disclosure from the IVI System's local maps data store. Using the definitions of the impact assessment table (Table \ref{Table_Impact_Assessment} \cite{Huang25A}), we evaluate \(i_{s} = 0\), \(i_{o} = 0\), and \(i_{f} = 0\), as there are no injuries or discernible operational and financial effects resulting from the disclosure of information. However, disclosure of personally identifiable information or passwords from the data store can result in privacy violations leading to abuses such as impersonation of the victim to perform actions with stolen identities. As such, the impact level for privacy is assessed as Medium and \(i_{p} = 10\). Using Equation \ref{eq:Equation_Impact_Sum_Normalised}, \(I_{\text{nsum}}\ = 0.0077\), which corresponds to a Low impact rating.

\begin{table}[H]
    \centering
    \caption{Impact rating example}
    \label{Table_Impact_Rating_Example}
    \begin{tabular}{|c|c|c|c|c|c|c|}
    \hline
    \textbf{Threat ID} & \textbf{$i_s$} & \textbf{$i_o$} & \textbf{$i_f$} & \textbf{$i_p$} & \textbf{Impact Score} & \textbf{Impact Rating} \\
    \hline
    9146 & 0 & 0 & 0 & 10 & 0.0077 & Low \\
    \hline
    \end{tabular}
\end{table}

\subsection{Step 5: Model attacks (cyber \& physical)}
In this step, we use an attack tree to model attacker goals and map out attack paths associated with both physical and cyber threat and attack vectors. Since the vehicle driver is within the car trust boundary (Figure \ref{fig:DFD_IVI}), we will assume the attacker personas and their goals as follows:

\begin{itemize}
    \item Persona A: A car thief interested in stealing sensitive data from the IVI system.
    \item Persona B: A rogue security researcher interested in manipulating vehicle functions that compromise the safety of the vehicle.
\end{itemize}

\begin{figure}[h]
    \centering
    \includegraphics[width=0.9\textwidth]{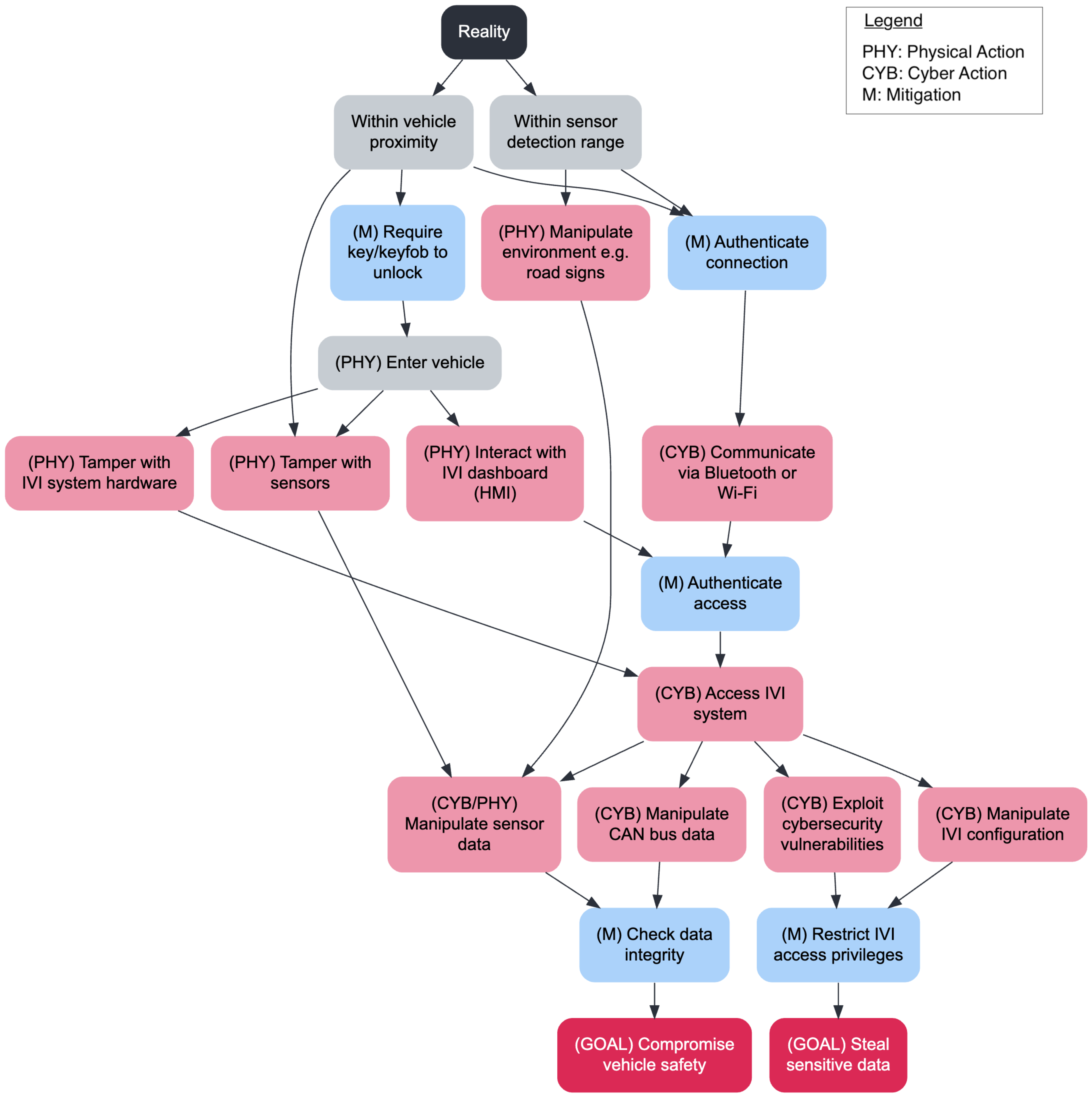}
    \caption{Attack tree for an In-Vehicle Infotainment (IVI) system, illustrating the combination of physical and cyber tactics to achieve the attacker's goals}
    \label{fig:Attack_Tree}
\end{figure}

The attack tree depicted in Figure \ref{fig:Attack_Tree} demonstrates how both personas could leverage a combination of physical and cyber tactics to achieve their objectives. One notable attack path included in the tree is the dirty road attack \cite{sato21}, where an attacker manipulates road signs (e.g. altering or obscuring traffic signals or directional signs) to mislead vehicle sensors. This physical manipulation could then be coupled with a cyber attack, such as exploiting a vulnerability in the IVI system to gain unauthorised access. By including both physical and cyber attack vectors, the attack tree underscores the importance of ACTISM's holistic approach to security modelling, addressing not only cyber but also physical attack surfaces that could be exploited by malicious actors.

\subsection{Step 6: Evaluate attack feasibility}

Next, we use CVSS (Common Vulnerability Scoring System) version 3.1 to analyse and compute attack feasibility. CVSS provides a standardised method for assessing the severity of vulnerabilities based on their impact and exploitability factors, as described below. We have chosen to use CVSS 3.1 instead of the more recent CVSS 4.0 because CVSS 3.1 is currently more widely adopted across our tools, processes, and industry standards. This ensures consistency in scoring and compatibility with existing vulnerability management systems.

\begin{itemize}
    \item \textbf{Impact Metrics}:
    \begin{itemize}
        \item Confidentiality (C): Impact on the confidentiality of data, such as driver information, vehicle location, and system diagnostics.
        \item Integrity (I): Impact on the integrity of automotive systems, including the accuracy and reliability of data and control signals within the vehicle.
        \item Availability (A): Impact on the availability of automotive functions and services, such as the ability to start the vehicle or use navigation systems.
    \end{itemize}
    
    \item \textbf{Exploitability Metrics}:
    \begin{itemize}
        \item Attack Vector (AV): How the vulnerability is accessed.
        \item Attack Complexity (AC): The complexity of the attack required to exploit the vulnerability.
        \item Privileges Required (PR): The level of privileges required to exploit the vulnerability.
        \item User Interaction (UI): The level of user interaction required to exploit the vulnerability.
    \end{itemize}
    
    \item \textbf{Impact Factor}:
    \begin{itemize}
        \item The Impact Factor depends on the scope of the vulnerability, which can be either Scope Unchanged (U) or Scope Changed (S). If the vulnerability affects other system components, the Scope is considered Changed (S), otherwise, it is Unchanged (U).
    \end{itemize}
\end{itemize}

In this experience report, instead of using the CVSS base score, which is common in the literature, we use the CVSS temporal score to incorporate and emphasise the temporal aspects of the vulnerability, such as exploit code maturity, remediation level, and report confidence. These are shown at Equations \ref{eq:Equation_CVSS_Base_Score} and \ref{eq:temporal_score}.

\begin{equation}
    \text{CVSS Base Score} = \text{Round up}\left( \frac{{0.6 \times \text{Impact} + 0.4 \times \text{Exploitability} - 1.5}}{{\text{Impact Factor} + 0.8}} \right)
    \label{eq:Equation_CVSS_Base_Score}
\end{equation}

\begin{equation}
    \begin{split}
    \text{CVSS Temporal Score} = \text{Round up}(\text{CVSS Base Score} \times \text{Exploit Code Maturity} \times \\
    \text{Remediation Level} \times \text{Report Confidence})
    \end{split}
    \label{eq:temporal_score}
\end{equation}

Next, the CVSS temporal scores for the identified threats are computed using the CVSS version 3.1 calculator (\url{https://nvd.nist.gov/vuln-metrics/cvss/v3-calculator}), then mapped to qualitative severity ratings using Table \ref{Table_CVSS_Rating_Scale} as referenced from the CVSS 3.1 Specification Document. The severity ratings, which we refer to as attack feasibility in this experience report, are tabulated in Table \ref{Table_Attack_Feasibility}.

\begin{table}[h]
    \centering
    \caption{Qualitative severity rating scale}
    \label{Table_CVSS_Rating_Scale}
    \begin{tabular}{|c|c|}
        \hline
        \textbf{CVSS Score} & \textbf{Severity Rating (Attack Feasibility)} \\
        \hline
        0.0 - 3.9 & Low \\
        \hline
        4.0 - 6.9 & Medium \\
        \hline
        7.0 - 8.9 & High \\
        \hline
        9.0 - 10.0 & Critical \\
        \hline
    \end{tabular}
    \vspace{-2.5em}
\end{table}

\begin{table}[h]
    \centering
    \caption{Attack feasibility based on CVSS version 3.1}
    \label{Table_Attack_Feasibility}
    \begin{tabular}{|c|c|c|c|c|c|c|c|c|c|c|c|c|c|c|}
    \hline
    \textbf{Threat ID} & \textbf{AV} & \textbf{AC} & \textbf{PR} & \textbf{UI} & \textbf{S} & \textbf{C} & \textbf{I} & \textbf{A} & \textbf{E} & \textbf{RL} & \textbf{RC} & \textbf{CVSS Score} & \textbf{Attack Feasibility} \\
    \hline
    9158 & A & L & N & N & C & N & L & H & P & U & U & 7.1 & High \\
    \hline
    9132 & A & H & N & N & C & N & H & H & U & U & U & 6.7 & Medium \\
    \hline
    9089 & P & L & N & R & U & H & H & H & U & U & U & 5.6 & Medium \\
    \hline
    9144, 9060 & A & L & L & R & U & H & H & H & U & U & U & 6.2 & Medium \\
    \hline
    9146 & P & H & N & R & U & H & L & N & U & U & U & 3.9 & Low \\
    \hline
    \end{tabular}
     \vspace{-2.5em}
\end{table}

\subsection{Step 7: Evaluate risks}
\begin{figure}[h]
    \centering
    \includegraphics[width=0.7\textwidth]{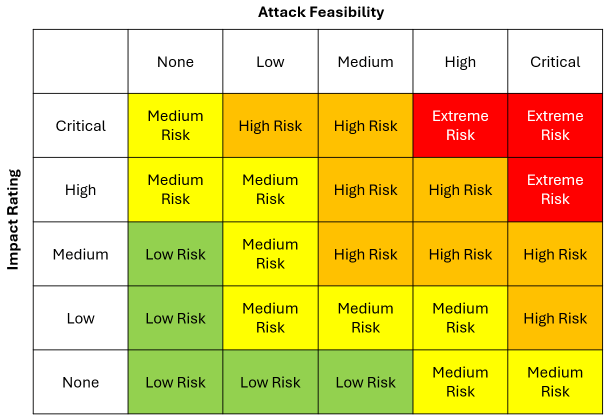}
    \caption{5x5 risk matrix}
    \label{fig:Risk_Matrix}
\end{figure}

In this step, we evaluate the cybersecurity risks of the IVI system using a 5x5 risk matrix illustrated in Figure \ref{fig:Risk_Matrix}. In the context of automotive impact assessment, where the stakes can be high and risks multifaceted, a 5x5 risk matrix provides the depth and granularity needed to accurately assess and manage risks, compared to a simpler 3x3 risk matrix. The risk ratings for the identified threats are tabulated in Table \ref{Table_Risk_Ratings}.

\begin{table}[h]
    \centering
    \caption{Risk ratings based on attack feasibility and impact ratings}
    \label{Table_Risk_Ratings}
    \resizebox{\textwidth}{!}{%
    \begin{tabular}{|c|p{4cm}|p{4cm}|c|c|c|c|}
    \hline
    \textbf{Threat ID} & \textbf{Description} & \textbf{Attack Method} & \textbf{Source} & \textbf{Attack Feasibility} & \textbf{Impact Rating} & \textbf{Risk Rating} \\
    \hline
    9132 &
    Cause a denial-of-service of the central gateway, affecting vehicle functions &
    Inject or flood invalid data to the central gateway causing it to crash or stop. &
    IVI System &
    High &
    Critical &
    Extreme \\
    \hline
    9159 &
    Cause gateway to misbehave, crash or stop, affecting vehicle functions &
    Exploit vulnerabilities, through outdated software or configuration weaknesses. &
    IVI System &
    Medium &
    High &
    High \\
    \hline
    9089 &
    Elevate privileges on the IVI System (via USB) &
    Exploit vulnerabilities, through outdated software or configuration weaknesses, via USB removable media. &
    USB &
    Medium &
    Low &
    Medium \\
    \hline
    9144, 9060 &
    Elevate privileges on the IVI System (via Bluetooth or Wi-Fi) &
    Exploit vulnerabilities, through outdated software or configuration weaknesses, via Bluetooth or Wi-Fi. &
    Bluetooth or Wi-Fi &
    Medium &
    Low &
    Medium \\
    \hline
    9146 &
    Access sensitive data e.g. passwords from IVI System Embedded Data Store &
    Reverse engineer the head unit firmware to find keys to access the IVI System Embedded Data Store. &
    Physical Access &
    Low &
    Low &
    Medium \\
    \hline
    \end{tabular}%
    }
\end{table}

These risk ratings serve as a guideline for the Original Equipment Manufacturer (OEM) of the vehicle or the manufacturer of the IVI system. They can use these ratings to prioritise the implementation of mitigation controls throughout the system life cycle. This approach ensures that adequate attention and resources are allocated to address the most critical cybersecurity concerns identified through this comprehensive analysis.

\subsection{Step 8: Review, update, and iterate}
In this final step, we continuously review the cybersecurity analysis and update the risk assessment based on new threat information, insights or changes in the automotive system environment. The updated risk assessment helps refine existing mitigation strategies and enhances security controls against emerging cybersecurity threats. This iterative cycle of review and improvement is crucial for maintaining the automotive system's long-term cybersecurity posture.

\section{Updating the Security Model After a Vulnerability Disclosure}
\label{sec:updating}

To demonstrate ACTISM's dynamic and iterative methodology, we simulate a response to the Tesla Electric Vehicle (EV) jail-break vulnerability disclosure \cite{Werling23}. While this vulnerability is specific to Tesla EVs, ACTISM is designed to be applicable to all automotive systems, including non-EVs, and their associated vulnerabilities.

Using technical details from the disclosure, we review the IVI system security model developed earlier and update the CVSS metrics as needed. In this case, no additional assets, DFD updates (Figure \ref{fig:DFD_IVI}), or attack tree revisions (Figure \ref{fig:Attack_Tree}) are required. Although new attack vectors, such as cyber-physical attacks like dirty road attacks \cite{sato21} or keyless entry systems \cite{Wouters21}, could be incorporated, we focus here on reassessing the CVSS metrics for Threat ID 9146, which aligns with the disclosed vulnerability. This process highlights ACTISM's adaptability in responding to real-world threats.

The CVSS metrics corresponding to Threat ID 9146 are updated as follows:

\begin{itemize}

    \item \textit{Scope (S)} From Unchanged (UC) to Changed (C), as successful exploitation of the vulnerability could grant an attacker full control over the IVI system, potentially enabling them to access and manipulate certain vehicle functions.
    
    \item \textit{Integrity (I)} From Low (L) to High (H), as successful exploitation of the vulnerability could grant an attacker root-level read and write access to the IVI file system.
    
    \item \textit{Availability (A)} From None (N) to High (H), as an attacker could deny access to resources in the IVI system.
    
    \item \textit{Exploit Code Maturity (E)} From Unproven (U) to Proof-of-Concept (P), in view of the disclosure from the security researchers.
    
    \item \textit{Remediation Level (RL)} From Unavailable (U) to Temporary Fix (T), as the AMD Trusted Platform Module (TPM) in affected IVI systems could still be susceptible to voltage glitching attacks.
    
    \item \textit{Report Confidence (RC)} From Unknown (U) to Confirmed (C), in view of the disclosure from the security researchers.
    
\end{itemize}

With these updated parameters, the attack feasibility increased from 2 (Low) to 3 (Medium), as shown in Table \ref{Table_Updated_Attack_Feasibility}.

\begin{table}[H]
    \centering
    \caption{Updated attack feasibility following vulnerability disclosure}
    \label{Table_Updated_Attack_Feasibility}    
    \resizebox{\textwidth}{!}{%
    \begin{tabular}{|c|c|c|c|c|c|c|c|c|c|c|c|c|c|c|}
    \hline
    \textbf{Threat ID} & \textbf{AV} & \textbf{AC} & \textbf{PR} & \textbf{UI} & \textbf{S} & \textbf{C} & \textbf{I} & \textbf{A} & \textbf{E} & \textbf{RL} & \textbf{RC} & \textbf{CVSS Score} & \textbf{Attack Feasibility} \\
    \hline
    9158 & A & L & N & N & C & N & L & H & P & U & U & 7.1 & High \\
    \hline
    9132 & A & H & N & N & C & N & H & H & U & U & U & 6.7 & Medium \\
    \hline
    9089 & P & L & N & R & U & H & H & H & U & U & U & 5.6 & Medium \\
    \hline
    9144, 9060 & A & L & L & R & U & H & H & H & U & U & U & 6.2 & Medium \\
    \hline
    \textbf{9146} & \textbf{P} & \textbf{H} & \textbf{N} & \textbf{R} & \textbf{C} & \textbf{H} & \textbf{H} & \textbf{H} & \textbf{P} & \textbf{T} & \textbf{C} & \textbf{6.4} & \textbf{Medium} \\
    \hline
    \end{tabular}%
    }
\end{table}

Besides attack feasibility, the impact assessment of Threat ID 9146 in the security model also needs updating. Using the definitions from the impact assessment table (Table \ref{Table_Impact_Assessment} \cite{Huang25A}), we estimate the impact on safety to increase from None to Low (\(i_{s} = 10\)), as there is now a possibility of light and moderate injuries, should the attacker access and manipulate certain vehicle functions via the vulnerability. We also expect the operational impact to increase from None to Low (\(i_{o} = 10\)), as there may be a degradation of primary vehicle functions. The financial impact is also expected to increase from None to Low (\(i_{f} = 10\)), as the financial damage from injury claims may be substantial but not threatening the existence of the manufacturer. Lastly, the privacy impact is expected to increase from Low to High (\(i_{p} = 100\)), as the privacy violation of multiple vehicle owners and drivers may lead to extensive media coverage and severe loss of market share. Using Equation \ref{eq:Equation_Impact_Sum_Normalised} to compute the updated normalised impact score, \(I_{\text{nsum}}\ = 0.1692\), which corresponds to a High impact rating (Table \ref{Table_Updated_Impact_Rating}).

\begin{table}[H]
    \centering
    \caption{Updated impact rating following vulnerability disclosure}
    \label{Table_Updated_Impact_Rating}
    \begin{tabular}{|c|c|c|c|c|c|c|}
    \hline
    \textbf{Threat ID} & \textbf{$i_s$} & \textbf{$i_o$} & \textbf{$i_f$} & \textbf{$i_p$} & \textbf{Updated Impact Score} & \textbf{Updated Impact Rating} \\
    \hline
    9146 & 10 & 10 & 10 & 100 & 0.1692 & High \\
    \hline
    \end{tabular}
\end{table}

With the updated attack feasibility (Medium) and impact rating (High), the overall risk rating for Threat ID 9146 increased from Medium to High, in accordance with the 5x5 risk matrix (Figure \ref{fig:Risk_Matrix}). This ability to perform timely updates to the risk assessment is crucial for maintaining robust cybersecurity in automotive systems. Ultimately, it empowers stakeholders to adopt additional precautions and improve the cybersecurity resilience of their vehicles throughout their life cycle.

\section{Practitioner Survey and Discussion}
\label{sec:surveydiscussion}

To assess the strengths of the ACTISM framework and identify areas for improvement, particularly in addressing emerging threats and enhancing usability in real-world automotive systems, feedback was gathered through a survey of 16 industry experts in Singapore. These experts included Chief Information Security Officers (CISOs) and cybersecurity practitioners specialising in security modelling and risk management. Representing a diverse range of industries across both private and public sectors, they provided a broad perspective on the evaluation of ACTISM. All respondents rated ACTISM as relevant, with a mean rating of 4.5 out of 5 based on a Likert scale, highlighting its potential as a robust framework for automotive cybersecurity. Furthermore, practitioners strongly agreed that ACTISM effectively integrates cyber and physical threat modelling (mean rating of 4.7 out of 5) and accurately identifies and models cyber-physical threats (mean rating of 4.1 out of 5). However, some respondents highlighted limitations in ACTISM, particularly its reliance on continuous threat intelligence updates to maintain effectiveness. They also noted challenges in implementing and using ACTISM compared to other security modelling frameworks. For automotive manufacturers with limited access to real-time cybersecurity threat intelligence feeds, as well as the absence of automation in the model workflow, these limitations could result in inconsistencies, false positives, and reduced operational efficiency over time.

To address these limitations and build on ACTISM's strengths, we propose several future directions, many of which align with the recommendations provided by the experts. A key area for improvement is the need for adequate real-world or simulated testing to enhance the framework's reliability. This would involve creating robust testing environments, including simulated attack scenarios, to evaluate ACTISM's performance under realistic conditions. Such testing would also help validate the accuracy of the model, particularly given the constantly evolving nature of threats and the potential for incomplete or inaccurate threat intelligence data.

Another recommendation from the experts is the integration of ACTISM with complementary frameworks such as the Exploit Prediction Scoring System (EPSS). Integrating EPSS, which predicts the likelihood of a vulnerability being exploited, alongside the Common Vulnerability Scoring System (CVSS), which measures the severity of a vulnerability, could enhance ACTISM's ability to prioritise and respond to threats effectively.

To further address the reliance on continuous threat intelligence updates, future work could focus on automating the model workflow. Establishing a comprehensive repository of vulnerability attributes, stored in a standardised format such as XML or JSON, would streamline the updating process and reduce the risk of inconsistencies. Additionally, formalising security models using Domain-Specific Languages (DSLs) like Automation Markup Language (AutomationML) and Meta Attack Language (MAL) could provide clearer descriptions of threats and vulnerabilities, improving the model's usability and accuracy.

Incorporating adversary tactics from the MITRE ATT\&CK knowledge base would further enhance ACTISM by offering insights into real-world attack strategies. This integration would enable the framework to better anticipate and mitigate sophisticated threats. Moreover, leveraging machine learning approaches to prioritise vulnerabilities based on their impact, utilising historical data and threat intelligence, could significantly improve the framework's predictive capabilities and overall effectiveness.

These future directions will ensure that automotive cybersecurity remains resilient and capable of addressing emerging challenges in an increasingly complex threat landscape.

\section{Conclusion}
\label{sec:conclusion}

The cybersecurity landscape for automotive systems is rapidly evolving, with attackers increasingly targeting vulnerabilities that threaten data integrity, functionality, and safety. The siloed application of traditional  security assessment methodologies using STRIDE and other frameworks often limits their ability to address the evolving and multi-faceted (cyber \& physical) nature of these threats. There is a clear knowledge gap in the ability to model and respond to security threats in a dynamic, iterative manner tailored specifically for automotive systems.

In response, we introduced ACTISM (Automotive Consequence-Driven and Threat-Informed Security Modelling), an integrated approach designed to enhance the resilience of automotive systems by dynamically updating their cybersecurity posture. ACTISM effectively responds to evolving cyber-physical threats and attacker tactics, ensuring that security models remain robust and relevant throughout a system's lifecycle.

We applied ACTISM to Tesla's In-Vehicle Infotainment system and showed that it can be practically used to model the dynamic nature of real-world cyber and physical threats in complex systems. 
Our experience report includes identifying key assets, conducting threat and attack modeling, and performing risk assessments for a complex system, in a dynamic, iterative manner. 
We also report the results of  practitioners' survey on the usefulness of ACTISM and its future directions.
In conclusion, ACTISM bridges the critical gap in automotive cybersecurity by offering a proactive and adaptable approach to security modelling. This framework not only addresses current challenges but also prepares for the future, ensuring the resilience and safety of automotive systems in an ever-evolving threat landscape.


%
%
%
\bibliographystyle{splncs04}
\bibliography{references}

\begin{table}
    \centering
    \caption{Threats generated by Microsoft Threat Modelling Tool}
    \label{Table_Threats}
    \resizebox{\textwidth}{!}{%
    \begin{tabular}{|c|c|p{5cm}|c|}
    \hline
    \textbf{Components} & \textbf{Threat ID} & \textbf{Threat Details} & \textbf{Threat Category} \\
    \hline
    In-Vehicle Informatics (IVI) System & 9042 & Adversary may perform a man-in-the-middle attack using hardware or software to access sensitive data. & Information Disclosure \\
    \hline
    Central gateway & 9132 & Adversary may inject or flood invalid data to the central gateway causing it to crash or stop. & Denial of Service \\
    \hline
    Central gateway & 9130 & Adversary may intercept/sniff data flowing from IVI to the central gateway, which may be used to attack other parts of the vehicle. & Information Disclosure \\
    \hline
    Central gateway & 9159 & Adversary may exploit vulnerabilities, through outdated software or configuration weaknesses. & Elevation of Privilege \\
    \hline
    USB & 9089 & Adversary may exploit vulnerabilities, through outdated software or configuration weaknesses through removable media. & Elevation of Privilege \\
    \hline
    Bluetooth & 9144 & Adversary may exploit vulnerabilities, through outdated software or configuration weaknesses through Bluetooth. & Elevation of Privilege \\
    \hline
    Wi-Fi & 9060 & Adversary may exploit vulnerabilities, through outdated software or configuration weaknesses through Wi-Fi. & Elevation of Privilege \\
    \hline
    GPS receiver & 9032 & Adversary may spoof GPS Signals and deliver malicious GPS data in order to cause drift off course. & Spoofing \\
    \hline
    GPS receiver & 9038 & Adversary may use a GPS jammer/send high power noise on the same frequency to prevent the GPS antenna signal from being received by the car. & Denial of Service \\
    \hline
    Human-Machine Interface (HMI) & 9050 & Adversary may flood the HMI (via the IVI) with invalid messages that prevents Advanced Driver Assistance Systems (ADAS) data being visualised. & Denial of Service \\
    \hline
    Local maps data store & 9146 & Adversary may reverse engineer the head unit firmware to find sensitive information. & Information Disclosure \\
    \hline
    IVI system embedded data store & 9122 & Adversary may use commands or actions provided in the IVI system to modify data in the IVI system embedded data store. & Tampering \\
    \hline
    IVI system embedded data store & 9124 & Adversary may reverse engineer the head unit firmware to find sensitive information. & Information Disclosure \\
    \hline
    \end{tabular}
    }
\end{table}

\begin{table}
    \centering
    \caption{Impact assessment table}
    \label{Table_Impact_Assessment}
    \resizebox{\textwidth}{!}{%
    \begin{tabular}{|p{3.4cm}|p{3.4cm}|p{3.4cm}|p{3.4cm}|c|c|}
    \hline
    \textbf{Safety} & \textbf{Operational} & \textbf{Financial} & \textbf{Privacy} & \textbf{Impact Level} & \textbf{Value} \\ 
    \hline
    No injury & No discernible effect & No discernible effect. No appreciable consequences & No discernible effects in relation to violations of privacy & None & 0 \\ 
    \hline
    Light and moderate injuries & Appearance item or audible noise (vehicle still operates, but does not conform, annoys more than 75\% of customers) & The financial damage remains tolerable to the organisation & Privacy violations of a particular stakeholder (e.g., vehicle owner, driver) which may not lead to abuses (e.g., impersonation of a victim to perform actions with stolen identities). Violation of legislations without appreciable consequences for business operations and finance (e.g., warning without any significant financial penalty, limited media coverage) for any stakeholder (e.g., OEM, fleet owner, driver) & Low & 1 \\ 
    \hline
    Severe and life-threatening injuries (survival probable) & Degradation of primary function (vehicle still operates, but at a reduced level of performance) & The resulting damage leads to substantial financial losses, but does not threaten the existence of the organisation & Privacy violations of a particular stakeholder (e.g., vehicle owner, driver) leading to abuses (e.g., impersonation of a victim to perform actions with stolen identities) and media coverage. Violation of legislations with potential consequences for business operations and finance (e.g., financial penalties, loss of market share, media coverage) & Medium & 10 \\ 
    \hline
    Life-threatening injuries (survival uncertain), fatal injuries & Potential failure mode affects safe vehicle operation without warning or involves non-compliance with government regulations & The financial damage threatens the existence of the organisation & Privacy violation of multiple stakeholders (e.g., fleet owners, multiple vehicle owners and multiple drivers) leading to abuses (e.g., impersonation of a victim to perform actions with stolen identities). Such a level of privacy violation may lead to extensive media coverage as well as severe consequences in terms of loss of market share, business operations, trust, reputation, and finance for OEMs and fleet owners. Violation of legislations (e.g., environmental, driver) causing significant consequences for business operations and finance (e.g., huge financial penalties, loss of market share) as well as extensive media coverage & High & 100 \\ 
    \hline
    \end{tabular}%
    }
\end{table}

\end{document}